
%
%
\hsize=6.0truein\vsize=8.5truein\voffset=0.25truein
\hoffset=0.1875truein
\tolerance=1000\hyphenpenalty=500
\def\monthintext{\ifcase\month\or January\or February\or
   March\or April\or May\or June\or July\or August\or
   September\or October\or November\or December\fi}

\def\monthdayandyear{\monthintext\space\number\day, \number\year}

\font\tenrm=cmr10 scaled \magstep1   \font\tenbf=cmbx10 scaled \magstep1
\font\sevenrm=cmr7 scaled \magstep1  
\font\fiverm=cmr5 scaled \magstep1   
\font\teni=cmmi10 scaled \magstep1   \font\tensy=cmsy10 scaled \magstep1
\font\seveni=cmmi7 scaled \magstep1  \font\sevensy=cmsy7 scaled \magstep1
\font\fivei=cmmi5 scaled \magstep1   \font\fivesy=cmsy5 scaled \magstep1

\font\tentt=cmtt10 scaled \magstep1
\font\tenit=cmti10 scaled \magstep1
\font\tensl=cmsl10 scaled \magstep1
\def\twelvepoint{\def\rm{\fam0\tenrm}
   \textfont0=\tenrm \scriptfont0=\sevenrm \scriptscriptfont0=\fiverm
   \textfont1=\teni  \scriptfont1=\seveni  \scriptscriptfont1=\fivei
   \textfont2=\tensy \scriptfont2=\sevensy \scriptscriptfont2=\fivesy
   \textfont\itfam=\tenit \def\it{\fam\itfam\tenit}
   \textfont\ttfam=\tentt \def\tt{\fam\ttfam\tentt}
   \textfont\bffam=\tenbf \def\bf{\fam\bffam\tenbf}
   \textfont\slfam=\tensl \def\sl{\fam\slfam\tensl} \rm
   \hfuzz=1pt\vfuzz=1pt
   \setbox\strutbox=\hbox{\vrule height 10.2pt depth 4.2pt width 0pt}
   \parindent=24pt\parskip=1.2pt plus 1.2pt
   \topskip=12pt\maxdepth=4.8pt\jot=3.6pt
   \normalbaselineskip=13.5pt\normallineskip=1.2pt
   \normallineskiplimit=0pt\normalbaselines
   \abovedisplayskip=13pt plus 3.6pt minus 5.8pt
   \belowdisplayskip=13pt plus 3.6pt minus 5.8pt
   \abovedisplayshortskip=-1.4pt plus 3.6pt
   \belowdisplayshortskip=13pt plus 3.6pt minus 3.6pt
   \topskip=12pt \splittopskip=12pt
   \scriptspace=0.6pt\nulldelimiterspace=1.44pt\delimitershortfall=6pt
   \thinmuskip=3.6mu\medmuskip=3.6mu plus 1.2mu minus 1.2mu
   \thickmuskip=4mu plus 2mu minus 1mu
   \smallskipamount=3.6pt plus 1.2pt minus 1.2pt
   \medskipamount=7.2pt plus 2.4pt minus 2.4pt
   \bigskipamount=14.4pt plus 4.8pt minus 4.8pt}
\twelvepoint

\def\letter{\parindent=0pt\parskip=\medskipamount\def\endmode{}
   \def\longindent{\parindent=3.25truein\obeylines\parskip=0pt}
   \def\letterhead{\null\vfil\begingroup
      \parindent=3.25truein\obeylines\parskip=0pt
      \def\endmode{\medskip\medskip\endgroup}}
   \def\date{\endmode\begingroup\parindent=3.25truein\obeylines\parskip=0pt
      \def\endmode{\medskip\medskip\endgroup}\monthdayandyear}
   \def\address{\endmode\begingroup
      \parindent=0pt\obeylines\parskip=0pt
      \def\endmode{\medskip\medskip\endgroup}}
   \def\salutation{\endmode\begingroup
      \parindent=0pt\obeylines\parskip=0pt\def\endmode{\medskip\endgroup}}
   \def\body{\endmode\begingroup\parskip=\medskipamount
      \def\endmode{\medskip\medskip\endgroup}}
   \def\closing{\endmode\begingroup\longindent
      \def\endmode{\endgroup}}
   \def\signed{\endmode\begingroup\longindent\vskip0.8truein
      \def\endmode{\endgroup}}
   \def\endofletter{\endmode \ifnum\pageno=1 \nopagenumbers\fi
      \vfil\vfil\eject}}

\def\preprint#1{ 
   \def\draft{\finishtitlepage{PRELIMINARY DRAFT:
\monthdayandyear}\writelabels%
      \headline={\sevenrm PRELIMINARY DRAFT: \monthdayandyear\hfil}}
   \def\date##1{\finishtitlepage{##1}}
   \font\titlerm=cmr10 scaled \magstep3
   \font\titlerms=cmr10 scaled \magstep1 
   \font\titlei=cmmi10 scaled \magstep3  
   \font\titleis=cmmi10 scaled \magstep1 
   \font\titlesy=cmsy10 scaled \magstep3 	
   \font\titlesys=cmsy10 scaled \magstep1  
   \font\titleit=cmti10 scaled \magstep3	
   \skewchar\titlei='177 \skewchar\titleis='177 
   \skewchar\titlesy='60 \skewchar\titlesys='60 
   \def\titlefont{\def\rm{\fam0\titlerm}
      \textfont0=\titlerm \scriptfont0=\titlerms 
      \textfont1=\titlei  \scriptfont1=\titleis  
      \textfont2=\titlesy \scriptfont2=\titlesys 
      \textfont\itfam=\titleit \def\it{\fam\itfam\titleit} \rm}
   \def\title##1{\vskip 0.9in plus 0.4in\centerline{\titlefont ##1}}
   \def\authorline##1{\vskip 0.9in plus 0.4in\centerline{\bf ##1}
      \vskip 0.12in plus 0.02in}
   \def\author##1##2##3{\vskip 0.9in plus 0.4in
      \centerline{{\bf ##1}\myfoot{##2}{##3}}\vskip 0.12in plus 0.02in}
   \def\addressline##1{\centerline{##1}}
   \def\abstract{\vskip 0.7in plus 0.35in\centerline{\bf Abstract}\smallskip}
   \def\finishtitlepage##1{\vskip 0.8in plus 0.4in
      \leftline{##1}\supereject\endgroup}
   \baselineskip=19pt plus 0.2pt minus 0.2pt \pageno=0
   \begingroup\nopagenumbers\parindent=0pt\baselineskip=13.5pt\rightline{#1}}

\def\nolabels{\def\eqnlabel##1{}\def\eqlabel##1{}\def\figlabel##1{}%
   \def\reflabel##1{}}
\def\writelabels{\def\eqnlabel##1{%
   {\escapechar=` \hfill\rlap{\hskip.11in\string##1}}}%
   \def\eqlabel##1{{\escapechar=` \rlap{\hskip.11in\string##1}}}%
   \def\figlabel##1{\noexpand\llap{\string\string\string##1\hskip.66in}}%
   \def\reflabel##1{\noexpand\llap{\string\string\string##1\hskip.37in}}}
\nolabels
\global\newcount\secno \global\secno=0
\global\newcount\meqno \global\meqno=1
\def\newsec#1{\global\advance\secno by1
   \xdef\secsym{\the\secno.}
   \global\meqno=1\bigbreak\medskip
   \noindent{\bf\the\secno. #1}\par\nobreak\smallskip\nobreak\noindent}
\xdef\secsym{}
\def\appendix#1#2{\global\meqno=1\xdef\secsym{\hbox{#1.}}\bigbreak\medskip
\noindent{\bf Appendix #1. #2}\par\nobreak\smallskip\nobreak\noindent}
\def\acknowledgements{\bigbreak\medskip\centerline{\bf
   Acknowledgements}\par\nobreak\smallskip\nobreak\noindent}
\def\eqnn#1{\xdef #1{(\secsym\the\meqno)}%
	\global\advance\meqno by1\eqnlabel#1}
\def\eqna#1{\xdef #1##1{\hbox{$(\secsym\the\meqno##1)$}}%
	\global\advance\meqno by1\eqnlabel{#1$\{\}$}}
\def\eqn#1#2{\xdef #1{(\secsym\the\meqno)}\global\advance\meqno by1%
	$$#2\eqno#1\eqlabel#1$$}
\def\myfoot#1#2{{\baselineskip=13.5pt plus 0.3pt\footnote{#1}{#2}}}
\global\newcount\ftno \global\ftno=1
\def\foot#1{{\baselineskip=13.5pt plus 0.3pt\footnote{$^{\the\ftno}$}{#1}}%
	\global\advance\ftno by1}
\global\newcount\refno \global\refno=1
\newwrite\rfile
\def\ref{[\the\refno]\nref}
\def\nref#1{\xdef#1{[\the\refno]}\ifnum\refno=1\immediate
   \openout\rfile=\jobname.aux\fi\global\advance\refno by1\chardef\wfile=\rfile
   \immediate\write\rfile{\noexpand\item{#1\ }\reflabel{#1}\pctsign}\findarg}
\def\findarg#1#{\begingroup\obeylines\newlinechar=`\^^M\passarg}
   {\obeylines\gdef\passarg#1{\writeline\relax #1^^M\hbox{}^^M}%
   \gdef\writeline#1^^M{\expandafter\toks0\expandafter{\striprelax #1}%
   \edef\next{\the\toks0}\ifx\next\null\let\next=\endgroup\else\ifx\next\empty%

\else\immediate\write\wfile{\the\toks0}\fi\let\next=\writeline\fi\next\relax}}
   {\catcode`\%=12\xdef\pctsign{
\def\striprelax#1{}
\def\semi{;\hfil\break}
\def\addref#1{\immediate\write\rfile{\noexpand\item{}#1}} 
\def\listrefs{\vfill\eject\immediate\closeout\rfile
   \centerline{{\bf References}}\medskip{\frenchspacing%
   \catcode`\@=11\escapechar=` %
   \input \jobname.aux\vfill\eject}\nonfrenchspacing}
\def\startrefs#1{\immediate\openout\rfile=refs.tmp\refno=#1}
\global\newcount\figno \global\figno=1
\newwrite\ffile
\def\fig{\the\figno\nfig}
\def\nfig#1{\xdef#1{\the\figno}\ifnum\figno=1\immediate
   \openout\ffile=\jobname.fig\fi\global\advance\figno by1\chardef\wfile=\ffile
   \immediate\write\ffile{\medskip\noexpand\item{Fig.\ #1:\ }%
   \figlabel{#1}\pctsign}\findarg}
\def\listfigs{\vfill\eject\immediate\closeout\ffile{\parindent48pt
   \baselineskip16.8pt\centerline{{\bf Figure Captions}}\medskip
   \escapechar=` \input \jobname.fig\vfill\eject}}

\def\noblackbox{\overfullrule=0pt}
\def\inv{^{\raise.18ex\hbox{${\scriptscriptstyle -}$}\kern-.06em 1}}
\def\dup{^{\vphantom{1}}}
\def\Dsl{\,\raise.18ex\hbox{/}\mkern-16.2mu D} 
\def\dsl{\raise.18ex\hbox{/}\kern-.68em\partial}
\def\slash#1{\raise.18ex\hbox{/}\kern-.68em #1}
\def\boxeqn#1{\vcenter{\vbox{\hrule\hbox{\vrule\kern3.6pt\vbox{\kern3.6pt
   \hbox{${\displaystyle #1}$}\kern3.6pt}\kern3.6pt\vrule}\hrule}}}
\def\mbox#1#2{\vcenter{\hrule \hbox{\vrule height#2.4in
   \kern#1.2in \vrule} \hrule}}  
\def\bar{\overline}\def\psibar{\bar\psi}
\def\e#1{{\rm e}^{\textstyle#1}}
\def\del{\partial}
\def\curly#1{{\hbox{{$\cal #1$}}}}
\def\curlyL{\hbox{{$\cal L$}}}
\def\vev#1{\langle #1 \rangle}
\def\lform{\hbox{$\sqcup$}\llap{\hbox{$\sqcap$}}}
\def\darr#1{\raise1.8ex\hbox{$\leftrightarrow$}\mkern-19.8mu #1}
\def\half{{\textstyle{1\over2}}} 
\def\roughly#1{\ \lower1.5ex\hbox{$\sim$}\mkern-22.8mu #1\,}
\def\MSbar{$\bar{{\rm MS}}$}
\hyphenation{di-men-sion di-men-sion-al di-men-sion-al-ly}
\def\ket#1{\vert #1 \rangle}\def\bra#1{\langle #1 \vert}
\def\Half{{1\over2}}
\def\d#1#2{d\mskip 1.5mu^{#1}\mkern-2mu{#2}\,}
\def\mev{\mathop{\rm Me\kern-0.1em V}\nolimits}
\def\gev{\mathop{\rm Ge\kern-0.1em V}\nolimits}
\def\alphaS{{\alpha_S}}\def\alphapi{{\alpha_S\over4\pi}}
\def\npb#1,#2,#3 {Nucl.\ Phys.\ {\bf B#1} (19#2) #3}
\def\prd#1,#2,#3 {Phys.\ Rev.\ D {\bf #1} (19#2) #3}
\def\plb#1,#2,#3 {\ifnum#1>170 Phys.\ Lett.\ B {\bf #1} (19#2) #3\else
 Phys.\ Lett.\ {\bf #1B} (19#2) #3\fi}
\def\sjnp#1,#2,#3 {Sov.\ J.\ Nucl.\ Phys.\ {\bf #1} (19#2) #3}
\def\yadfiz#1,#2,#3 {Yad.\ Fiz.\ {\bf #1} (19#2) #3}

\def\curlyO{\curly O}
\def\onezero{({\bf 1}\thinspace\thinspace{\bf 0})}
\def\bdagger{b^\dagger}
\def\fsubB{$f_B$}
\def\zerohat{{\bf \hat 0}}
\def\muhat{\hat \mu}
\noblackbox
\preprint{hep-ph@xxx/9203221}\rightline{UCLA/92/TEP/9}\rightline{MCGILL/92--11}
\vskip 0.3in plus 0.1in
\title{\titlefont Improved Heavy Quark Effective Theory Currents}
\vskip 0.85in plus 0.4in
\centerline{\bf Oscar F. Hern\'andez}
\vskip 0.12in plus 0.02in
\addressline{Department of Physics}
\addressline{McGill University}
\addressline{Ernest Rutherford Physics Building}
\addressline{Montr\'eal, Qu\'e., Canada H3A 2T8}
\vskip 0.2in plus 0.02in
\centerline{\bf Brian R. Hill}
\vskip 0.12in plus 0.02in
\addressline{Department of Physics}
\addressline{University of California}
\addressline{Los Angeles, CA~~90024}
\vskip 0.1in plus 0.02in
\abstract
It is hoped that the accuracy of a variety of lattice calculations
will be improved by perturbatively eliminating effects proportional to the
lattice spacing.  In this paper, we apply this improvement program to the heavy
quark effective theory currents which cause a heavy quark to decay to a light
quark, and renormalize the resulting operators to order $\alphaS$.
We find a small decrease in the amount that the operator needs to be
renormalized, relative to the unimproved case.
\vskip 0.1in plus 0.02in
\date{3/92}
\newsec{Introduction}%
It is hoped that the accuracy of a variety of lattice
calculations will be improved by the elimination of effects proportional to
powers of the lattice spacing, $a$~\ref\program{K. Symanzik, Nucl. Phys. B
{\bf 226} (1983) 187.}\ref\LandW{
M. L\"uscher and P. Weisz, Commun. Math. Phys. {\bf 97} (1985) 59.}.
This improvement program can be implemented
order by order in $a$ and the strong coupling $g^2$.  Calculationally, it is
relatively easy to perform the leading order in either of these expansions. For
matrix elements of vector currents, both types of leading order corrections are
thought to be in
the~20 to~30\% range~\ref\Tallahassee{E. Gabrielli {\it et al},
in {\sl Lattice 90,} ed. by U. M. Heller {\it
et al,} Nucl. Phys. B (Proc. Suppl.) {\bf 20} (1991) 448\semi G.
Heatlie {\it et al,} \npb352,91,266 .},
so it is at the same time important to
include them, and reasonable to stop at this order.

Much of the application of the improvement program to hadronic matrix elements
has been to the order $a$
improvement of the Wilson fermion action and operators
\ref\NNN{H. W. Hamber and C. M. Wu, \plb133,83,351 ; \plb136,84,255 .}--%
\nref\gab{E. Gabrielli {\it et al,} \npb362,91,475 .}%
\ref\SandW{B. Sheikholeslami and R. Wohlert, \npb259,85,572 .}.
Alternative methods for measuring matrix elements involving
quarks that are heavy relative to the QCD scale
have been proposed~\ref\Etalk{E. Eichten, in {\sl Field
Theory on the Lattice,} edited by A.~Billoire {\it et al.}, Nucl.
Phys. B (Proc. Suppl.) {\bf 4} (1988) 170.}\ref\LandTtalk{G. P.
Lepage and B. A. Thacker, {\it ibid}, p. 199.}.
The action and operators in this method can be thought of
as discretizations of the heavy quark effective field theory action and
operators~\LandTtalk
\ref\eft{W. E. Caswell and G. P. Lepage, \plb167,86,437\semi
H. D. Politzer and M. B. Wise, \plb208,88,504 \semi
E. Eichten and B. Hill, \plb234,90,511 \semi
B. Grinstein, \npb339,90,253 \semi
H.~Georgi, \plb240,90,447 \semi
M. J. Dugan, M. Golden, and B. Grinstein, HUTP/91--A045, to appear in Phys.
Lett. B.}.
In this paper, we study the application of the improvement program to this
action and the currents which cause a heavy quark to decay to a light quark.

The outline of the paper is as follows.  In the next section, we discuss
improvement of the heavy quark action, and in section three, we discuss
improvement of the aforementioned currents.  As is generally the case, the
improved lattice currents need to be related  to their continuum counterparts.
This is done perturbatively in $g^2$ where the scale is set by the lattice
spacing.  Although quite different in motivation, this perturbative calculation
is similar to the renormalization of other discretizations of the heavy quark
effective theory currents performed in
reference~\ref\fbsplit{O. F. Hern\'andez and B. R. Hill, UCLA/91/TEP/51, to
appear in Phys. Lett.~B.}. That calculation and the present one rely on the
framework and results of reference~\ref\fb{E. Eichten and B. Hill,
\plb240,90,193.}, which are summarized in section~four before the new
perturbative results are presented. In section five we give our conclusions.
\newsec{Improvement of the Heavy Quark Action}%
In this section, we argue that
under two criteria of improvement, there is no need to modify the heavy quark
action.  The arguments are tree level, but this is adequate for the leading
order improvement discussed in the introduction.

Roughly, the on-shell improvement condition formulated by L\"uscher and
Weisz~\LandW\ is
that spectral quantities such as the location of single-particle poles
should not be corrected by terms proportional to powers of the lattice spacing.
Examining the momentum space propagator for the heavy quark action in
the continuum~\eft,
\eqn\contprop{{1\over p_0+i\epsilon},}
and comparing it with its lattice counterpart~\fb,
\eqn\latticeprop{{1\over{-i}(e^{ip_0a}-1)/a+i\epsilon},}
we see that although the propagators differ at order $a$, the
pole is located at $p_0=0$ to all orders in $a$.  We immediately conclude that
the action has no need of improvement under the on-shell improvement condition,
to all orders in $a$ at tree level.

A second argument yielding this conclusion is to compare
the heavy quark propagator in position space in an external gauge field
on the lattice to the same thing in the
continuum.\myfoot{$^\dagger$}{We thank Estia
Eichten for discussions leading to this argument.}
In the continuum, the propagator from $x$ to $y$
is~\ref\olderpapers{J. M. Cornwall and G. Tiktopoulos, \prd15,77,2937 \semi
E. Eichten and F. Feinberg, \prd23,81,2724 .},
\eqn\poscontprop{-i\delta^3({\bf y}-{\bf x})\theta(y_0-x_0)
P\exp{-ig\int_{x_0}^{y_0}dt\thinspace A_0(t,{\bf x})}.}
The discretization of the action with a nearest neighbor,
one-sided time derivative follows from this propagator~\fb.
On the lattice, the heavy quark propagator from~$m$ to~$n$ in a
background field is~\Etalk\LandTtalk,
\eqn\poslattprop{{-i\over a^3}\delta_{\bf nm}\theta(n_0-m_0)
U_0(n-\zerohat)^\dagger
U_0(n-2\zerohat)^\dagger
\cdots
U_0(m)^\dagger.}
In this expression, $\delta_{\bf nm}$ is the three-dimensional
Kronecker $\delta$--function,
$\theta(n_0-m_0)$ is $1$ if $n_0\ge m_0$ and $0$ otherwise, and $U_0(m)$ is
the lattice gauge link in the time direction at site $m$.

If the correspondence between continuum and lattice gauge fields is~\LandW,
\eqn\correspondence{U_0(n)=
P\exp{iga\int_0^1dt\thinspace A_0((n_0+1-t)a,{\bf n}a)},}
then the lattice propagator perfectly reproduces its continuum counterpart.
If one modifies the heavy quark action (by choosing any other
discretization of the time derivative)
the position space lattice propagator~\poslattprop\ is multiplied by a
$c$-number factor depending only on $n_0{-}m_0$ but is otherwise unchanged
(as long as the discretization does not include spatial links).  With the
correspondence~\correspondence, there is clearly no closer approximation
to the continuum propagator~\poscontprop\ than the propagator~\poslattprop\
obtained from the action currently in use.
\newsec{Improved Heavy-Light Currents}%
Since the lattice heavy quark action is not in need of improvement, we review
the situation for the light quark action.  Expressions for improved
lattice heavy-light currents quickly follow from this discussion.

The improved Wilson fermion action proposed by Sheikholeslami and
Wohlert~\SandW\ is
obtained from an improved action with next-to-nearest neighbor couplings~\NNN\
by a change of variables.  The resulting action is the Wilson action
plus an additional piece,
\eqn\DSI{\Delta S_I=-ia^4{ar\over4}\sum_{n\mu\nu}
\bar q(n)\sigma_{\mu\nu}gP_{\mu\nu}(n)q(n).}
$P_{\mu\nu}(n)$ is the sum of plaquettes defined in reference~\gab\
and goes to $F_{\mu\nu}(n)$
in the continuum limit, while $\sigma_{\mu\nu}=[\gamma_\mu,\gamma_\nu]/2$.
The action has the advantage that it only contains nearest neighbor couplings.
One can most easily obtain improved operators for use with this action by
starting with local bilinears and making the replacement~\Tallahassee\SandW,
\eqn\transformation{q(n)\rightarrow q(n)-{r\over2}\sum_\mu\gamma_\mu
\left[U_\mu(x)q(n+\mu)-U_\mu(n-\muhat)^\dagger q(n-\muhat)\right].}
A similar replacement is made for $\bar q(n).$

Actually, a two-parameter family of transformations, all of
which yield operators improved to order $a$, can be obtained by
using the equations of motion for the light and heavy quark.  The
effect of applying the equation of motion to the light quark field has
been discussed in reference \gab\ and the effect of applying the
equation of motion to the heavy quark field has been studied under
the guise of temporally split operators~\fbsplit.  In either case,
the operator renormalization is changed by an amount which comes
from the self-energy.  We will not consider this generalization further.

{\baselineskip=18pt plus 1pt
The most general heavy-light bilinear in the full continuum theory is,
\eqn\J{J(x)=\bar b(x) {\bf \Gamma} q(x).}
Here ${\bf \Gamma}$ is any Dirac matrix, and $q$ is the light quark field.
In the heavy quark effective theory, the corresponding operator~\J\ is~\eft,
\eqn\cont{\bdagger(x)\onezero{\bf \Gamma} q(x).}
In a Dirac basis, the two-by-four matrix preceding ${\bf \Gamma}$
takes the form $\onezero$ and projects onto the upper two rows of
${\bf \Gamma}.$  Applying the above recipe to the local operator
$b^\dagger(n)\onezero{\bf \Gamma q(n)}$ we are led to consider,
\eqn\improvedcurrent{b^\dagger(n)\onezero{\bf \Gamma} q(n)
-{r\over2}b^\dagger(n)\onezero{\bf \Gamma}\sum_\mu\gamma_\mu
\left[U_\mu(x)q(n+\mu)-U_\mu(n-\muhat)^\dagger q(n-\muhat)\right].}
The first term in this expression is the usual local current used to determine
hadronic matrix elements.  The calculations necessary to renormalize this part
of the current were mostly done in previous papers~\fb\ref\blp{Ph. Boucaud,
C. L. Lin, and O. P\`ene, \prd40,89,1529 ; \prd41,90,3541(E) .},
however it receives
additional contributions coming from the new term in the action \DSI.  The
remainder of the current,  which is naively order $a,$ will also affect the
renormalization of the current when it appears in loop diagrams with loop
momenta of order $a^{-1}$. We now turn to the perturbative renormalization of
the improved heavy-light current with the improved Wilson action of
Sheikholeslami and Wohlert.
\newsec{Renormalization of the Improved Current}%
The lattice renormalization of
the heavy-light current \improvedcurrent\ gives the ratio of the lattice
operator to its counterpart in the continuum theory.  We can divide the
diagrams that contribute to this ratio into two parts.  Those that give
heavy and light quark wave function renormalization, and the 1PI
vertex correction diagrams.

The difference between wave function renormalization of the heavy quark on the
lattice and in the continuum is $g^2/(12\pi^2)$ times a constant $e=4.53$
computed in references~\fb\ and~\blp.\myfoot{$^\dagger$}{Here
we are using the reduced value of $e$ which is appropriate
if one fits correlation functions containing the propagator
\poslattprop\ to $A\e{-B(n_0{-}m_0)a}$~\fbsplit\fb\blp.}
The corresponding light quark wave function renormalization constant was
calculated in \gab.
The results for the~self~energy-graphs~which\eject}
we will need are given in terms of $\Delta_{\Sigma_1}$, defined in
reference \gab\ (when consulting their expressions for $\Delta_{\Sigma_1},$
note that we have taken $F_{0001}=1.31$~\ref\fzzzo{A. Gonzales-Arroyo and C. P.
Korthals-Altes, \npb205,83,46. }).

We now turn to the vertex correction diagrams of figures
\fig\veri{The vertex correction diagrams resulting from improving the
light quark operator} and \fig\verii{The vertex correction diagrams
resulting from improving the action}.  The techniques we use to
evaluate these diagrams have been discussed in detail in a number of
references~\fbsplit\fb, and will not be reviewed here.

At zero external momentum, \veri (a) vanishes.  As noted below
equation~\improvedcurrent, the operator has two pieces: one which is the
local unimproved operator, and an additional piece coming from improvement.
Thus the contribution of \veri (b) can be split up into two parts
which we will call the ``unimproved'' and ``improved'' contributions here
and in the following paragraph.  The ``unimproved'' contribution is~\fb\blp,
\eqn\verib{
{g^2\over 12 \pi^2} \big( d_1 + G d_2  \big)
.}
We have defined the
$c$-number $G$ by $G{\bf \Gamma}=\gamma_0{\bf \Gamma}\gamma_0$.
The analytical expressions for $d_1$ and $d_2$ can be found in reference~\fb.
We tabulate the constant
$\Delta_{\Sigma_1}$ and the constants $d_1$ and $d_2$ in Table~1
for several values of the Wilson mass parameter $r$.
Errors in Table~1 and~2 are at most \curlyO(1)\ in the last decimal place.
\topinsert
\vbox{
\def\tablerule{\noalign{\hrule}}
$$\vbox{\offinterlineskip\tabskip=0em
	\halign to 4in{\vrule #\tabskip=1em plus5em&
		\strut\hfil$#$\hfil&\vrule #&
      \hfil$#$\hfil&\vrule #&
      \hfil$#$\hfil&\vrule #&
      \hfil$#$\hfil&\tabskip=0em\vrule #\cr
		\tablerule
		&r&&d_1&&d_2&&\Delta_{\Sigma_1}&\cr
		\tablerule
		&1.00&&5.46&&-7.22&&-9.21&\cr
		\tablerule
		&0.75&&5.76&&-7.23&&-8.62&\cr
		\tablerule
		&0.50&&6.30&&-7.00&&-7.80&\cr
		\tablerule
		&0.25&&7.37&&-5.72&&-6.73&\cr
		\tablerule
		&0.00&&8.79&&\phantom{-}0.00&&-6.04&\cr
		\tablerule}}
$$
\centerline{Table 1. Previously Computed $r$-dependent
Quantities \gab\fb\blp.}
}
\endinsert

The ``improved'' contribution from diagram \veri (b) can be combined
with the contribution from the other three diagrams in figure \veri.
In order to compactly quote the analytic expressions for this result, we
use the notation of references
\ref\MandZ{G. Martinelli and Y.-C. Zhang, \plb123,83,433 .}
and~\ref\fhh{J. M. Flynn, O. F.  Hern\'andez, and B. R. Hill,
\prd43,91,3709 .}\
for the following commonly occurring combinations:
\def\summu{\sum\nolimits_\mu}
\eqn\Deltas{\eqalign{\Delta_1=\summu\sin^2{l_\mu\over2},\quad &
\Delta_4=\summu\sin^2{l_\mu}, \cr
\Delta_2=\Delta_4+4r^2\Delta_1^2 , \quad &
\Delta_5=\summu\sin^2{l_\mu\over2}\sin^2{l_\mu}.\cr
}}
Additional combinations, $\Delta_1^{(3)},$ $\Delta_2^{(3)},$ and
$\Delta_4^{(3)},$ are the same as above except the sums on $\mu$ run only
from~$1$ to~$3$.
Given these definitions the total ``improved'' contribution from
figure~\veri\ to the vertex renormalization is,
\eqn\totalveri{
 {g^2\over 12 \pi^2} ( \Delta d_1 + G \Delta d_2),
}
where,
\eqn\ij{
\Delta d_1 ={-3r^2 \over 16}\int {d^4 l \over \pi^2} {4-\Delta_1 \over
\Delta_2},
\quad \Delta d_2={-r^3 \over 2}\int {d^3 l \over \pi} {\Delta^{(3)}_1 \over
\Delta^{(3)}_2}.
}

We now turn to the contributions depicted in figure~\verii\ which are
due to the additional term in the
action~\DSI.  In these figures, the insertion of this term
is denoted with a cross.  A tadpole
diagram which vanishes has not been depicted.  The result
for the two diagrams depicted is,
\eqn\totalverii{
{g^2\over 12 \pi^2} ( \Delta d'_1 + G \Delta d'_2 ),
}
where,
\eqn\ijkl{
\Delta d'_1 ={r^2 \over 4}\int {d^4 l \over \pi^2}
{\Delta_4(3-\Delta_1)+\Delta_5 \over
4\Delta_1\Delta_2},
\quad \Delta d'_2 ={r \over 2}\int {d^3 l \over \pi}
{\Delta^{(3)}_4(1+r^2\Delta^{(3)}_1) \over
4\Delta^{(3)}_1\Delta^{(3)}_2}.
}

\topinsert
\def\tablerule{\noalign{\hrule}}
\vbox{
$$\vbox{\offinterlineskip\tabskip=0em
	\halign to 6in{\vrule #\tabskip=1em plus5em&
		\strut\hfil$#$\hfil&\vrule #&
      \hfil$#$\hfil&\vrule #&
      \hfil$#$\hfil&\vrule #&
      \hfil$#$\hfil&\vrule #&
      \hfil$#$\hfil&\tabskip=0em\vrule #\cr
		\tablerule
&r&&\Delta d_1&&\Delta d_2&&\Delta d_1'&&\Delta d_2'&\cr
		\tablerule
		&1.00&&-6.64&&-5.84&&3.43&&6.16&\cr
		\tablerule
		&0.75&&-5.30&&-3.75&&2.53&&4.94&\cr
		\tablerule
		&0.50&&-3.61&&-1.87&&1.55&&3.66&\cr
		\tablerule
		&0.25&&-1.52&&-0.45&&0.55&&2.15&\cr
		\tablerule
		&0.00&&\phantom{-}0.00&&\phantom{-}0.00&&0.00&&0.00&\cr
		\tablerule}}
$$
\centerline{Table 2. Changes to $d_1$ and $d_2$ as a function of $r$.}
}
\endinsert

Combining the various results in this section, we find that the
ratio of the lattice to continuum operators is,
\eqn\ratiolec{
1+{g^2\over12\pi^2}\left[
(d_1+\Delta d_1 + \Delta d'_1)
+(d_2+\Delta d_2 + \Delta d'_2)G
+{1\over2}e-{1\over2}\Delta_{\Sigma_1}-1 \right]\!.}
The dependance on $\mu a$ has been eliminated by setting $\mu=1/a$.
\newsec{Conclusions}%
We illustrate the use of the results of the previous section for the case of
most
interest, the current which determines the $B$ meson decay constant, $f_B$.  In
that case, for reasonable values of the input parameters, the ratio of the
continuum effective theory to full theory bilinears is numerically
$0.98$~\eft\blp.
We need to multiply this by the ratio given in eq.~\ratiolec.
For ${\bf \Gamma}=\gamma_0\gamma_5,$ the constant $G$
(which appears in \ratiolec) is~$-1$. We take $g^2=1.8$ which is appropriate
for effects arising from the scale $\pi/a$ with $1/a=2\gev$ as argued
in~\ref\LandM{G. P.
Lepage and P. B. Mackenzie, in {\sl Lattice 90,} ed. by U. M. Heller
{\it et al,} Nucl. Phys. B (Proc. Suppl.) {\bf 20} (1991) 173.}.
Taking values from tables~1 and~2 with $r=1.00$,
equation~\ratiolec\ gives 
1.23 and the product of the two ratios is 1.20 (as compared to 1.28 in the
unimproved case).  To obtain the physical value of $f_B,$ one divides
the lattice results for the improved current by this number.

The operator we have renormalized is corrected both to order $g^2$ and to
order~$a$.   As noted in the introduction both of these corrections are thought
to be in the 20 to 30\% range.  Analytically, the next perturbative
corrections are proportional to $g^2 a,$ $g^4,$ or $a^2$ times
powers of $g^2\ln a$~\Tallahassee, and there is numerical evidence
that these further corrections are at the few per cent level for
currents made from two Wilson fermion fields~\Tallahassee.  Thus it is hoped
that the improved currents renormalized here will lead to a considerably
more precise lattice determination of $f_B$ and other heavy quark
matrix elements.
\vskip 0.2in
{\it Note Added.} While preparing this manuscript we learned of unpublished
work on this  subject~\ref\asquared{G. Martinelli and G. C. Rossi,
unpublished.}\ref\BandP{A. Borrelli and C. Pittori, unpublished.} referenced in
a paper on lattice measurements of several quantities using an improved
action~\ref\results{G. Martinelli, C. T. Sachrajda, G. Salina, and A. Vladikas,
SHEP 91/92--6.}.   Martinelli and Rossi~\asquared\ argue that
it is unnecessary to change the heavy quark action up to order $a^2.$
This is consistent with our conclusion in section~two that the heavy quark
action is not in need of improvement at any order in $a$.  The result of
Borrelli and Pittori~\BandP\ cited in reference~\results\ is consistent with
our factor above if one takes $g^2=1.0$ rather than~$1.8$.
\acknowledgements
OFH was supported in part by the National Science and Engineering Research
Council of Canada, and les Fonds FCAR du Qu\'ebec.
BRH was supported in part by the Department of Energy under Contract No.
DE--AT03--88ER 40383 Mod~A006--Task~C.
\baselineskip=16pt\listrefs\listfigs\bye